\documentclass[aps,prl
,twocolumn,floatfix,superscriptaddress,showpacs]{revtex4-1}
\usepackage{amsmath}
\usepackage[dvips]{graphicx}
\usepackage{ulem}
\begin{document}

\newcommand{\ra}{$t_\downarrow/t_\uparrow$}
\newcommand{\tup}{$t_\uparrow$}
\newcommand{\tdw}{$t_\downarrow$}
\newcommand{\up}{$\uparrow$}
\newcommand{\dw}{$\downarrow$}
\newcommand{\tn}{$T_N$}
\newcommand{\docc}{$\langle n_\uparrow n_\downarrow\rangle$}
\newcommand{\dav}{$\langle d\rangle$}
\newcommand{\cv}{$C_v$}

\title{Orbital selective crossover and Mott transitions in an asymmetric Hubbard model
of cold atoms in optical lattices}
\author{E. A. Winograd}
\affiliation{Laboratoire de Physique des Solides, CNRS-UMR8502, Universit\'e de Paris-Sud,
Orsay 91405, France.}
\author{R. Chitra}
\affiliation{Laboratoire de Physique Theorique de la Mati\`ere 
Condense\'e, UMR 7600, Universit\'e de Pierre et Marie Curie, Jussieu, Paris-75005, France.}
\author{M. J. Rozenberg}
\affiliation{Laboratoire de Physique des Solides, CNRS-UMR8502, Universit\'e de Paris-Sud,
Orsay 91405, France.}
\affiliation{Departamento de F\'{\i}sica, FCEN, Universidad de Buenos Aires,
Ciudad Universitaria Pab.I, (1428) Buenos Aires, Argentina.}

\date{\today}

\begin{abstract}
We study  the asymmetric Hubbard model at half-filling
as a generic model  to describe the physics of two species of
repulsively interacting   fermionic  cold atoms in optical lattices.
We use Dynamical Mean Field Theory
to obtain the paramagnetic  phase diagram of the model as function of
temperature, interaction strength and  hopping asymmetry.
A Mott transition with a region of two coexistent solutions is found
for all nonzero values of the hopping asymmetry.
At low temperatures the metallic phase is a heavy Fermi-liquid, qualitatively analogous to  the 
Fermi liquid state of the symmetric Hubbard model. Above a coherence
temperature, an orbital-selective crossover takes place, wherein
one fermionic species effectively localizes, and the resulting bad metallic state
resembles the non-Fermi liquid state of the Falicov-Kimball model.
We compute observables relevant to cold atom systems such as the
double occupation, the specific heat
and entropy and characterize their behavior in the different phases.
\end{abstract}

\maketitle

Quantum degenerate fermions can now be loaded onto optical lattices to 
recreate model hamiltonians with easily tunable physical parameters  \cite{rmpcoldatoms}.
Two fermionic atomic species are minimally required to create a spin-1/2 model, where
each type of atom is associated to a different projection of the spin.  
This can be achieved by either loading the optical lattice 
with  atoms with two different hyperfine spin states or 
by two different atom species
\cite{rmpcoldatoms,giamarchi,sarma,dao,lda}.
Remarkable aspects
of the cold atom systems include  the variety of lattice-types that can be realized, and
the large range of interaction strengths that can be accessed, that may be up to several times the
fermion bandwidth\cite{esslinger-review}. 
A lot of recent work has focused on the fermionic  Hubbard model in 3D optical lattices, which is
the paradigm of a strongly correlated fermionic system. 
For instance, a BCS-BEC crossover was observed in
the attractive case using $^6Li$ atoms \cite{chin,rmpfg}, and also a Mott insulating state
was reported in the repulsive case at half-filling using $^{40}K$ atoms\cite{jordens,science}.  
Although the exact realization of the Hubbard model  may be achieved in cold atom 
systems, the generic situation, involving two different fermionic atomic
species, will not exactly fulfill the requirement of spin symmetry. Thus, the 
model  typically realized in an optical lattice would be an Asymmetric Hubbard model (AHM), 
which interpolates between the Hubbard model and the spinless Falicov-Kimball model (FKM), 
with or without population imbalance.  
This model, in the case of attractive interactions, has been considered in recent theoretical 
investigations that focused on the possibility of superconducting
phases \cite{giamarchi,dao}.
The study of the  AHM with repulsive interactions  is therefore  relevant and  may provide 
needed guidance for current experimental investigations. 

In this paper, we focus on the interesting case of the half filled AHM 
with no population imbalance and the associated  paramagnetic  Mott
metal-insulator transition.
For simplicity, we neglect the harmonic
potential in our calculations, thus our results apply to the bulk of the optical lattice, 
away from the edges of the trap.
We study the model within Dynamical
Mean Field Theory (DMFT), a method which has  significantly
advanced our understanding of  the Mott transition phenomenon \cite{bible,fk}. 
Here, we focus on paramagnetic
states which may  be realized in frustrated optical lattices.
Our work shows that, akin to the Hubbard case,
there is a region of coexistent solutions defined by two Mott-transitions at critical values
$U_{c1}$ and $U_{c2}$ \cite{bible}. This region exists within the entire asymmetric regime, except right at
the FKM end-point. While this observation may suggest that the AHM qualitatively
behaves like the ordinary Hubbard model, we find that the actual situation is more subtle and
cannot be described in terms of a simple interpolation between the two known limiting cases. 
In fact, in the metallic phase at low temperatures, we find
a heavy-mass Fermi liquid state  which is qualitatively  similar to
that of the spin-symmetric Hubbard model \cite{bible}.
However, above a coherence temperature $T_{coh}$, instead of the
standard incoherent metal seen in the Hubbard case,  an orbital-selective
crossover occurs, where the solution evolves to closely resemble the  bad metallic 
state of the FKM. Thus, surprisingly, the behavior of the AHM embodies
features of both, the Hubbard and the FKM.
We also compute the  double occupancy which is experimentally
accessible \cite{jordens}, and the specific heat and entropy which are useful for the
estimation of the temperature of experimental systems \cite{science}.

The Asymmetric Hubbard model Hamiltonian reads,
\begin{equation}
\label{eq:ahm}
H=-\sum_{<ij>;\sigma=\uparrow,\downarrow} t_\sigma c_{i\sigma}^\dag c_{i\sigma}^{\phantom{\dag}}
+U\sum_i (n_{i\uparrow}-1/2)(n_{i\downarrow}-1/2)
\end{equation}
$c,c^\dag$ are the fermion creation and annihilation 
operators and the  indices  $(i,j)$ and $\sigma$ label the  lattice sites and the two species of  fermions, respectively.
Defining $r \equiv t_\downarrow /  t_\uparrow $, the AHM interpolates
between  the FKM at $r=0$ (where one of the 
fermion species  is immobile)  and  the
 the spin-symmetric Hubbard model at $r=1$. 
These two limits were studied extensively  within the framework of 
DMFT (cf. see reviews of Refs. \onlinecite{bible} and \onlinecite{fk}).
We adopt a bounded semi-circular density of states  for both fermionic bands with
the half-bandwidths $D_\sigma = 2 t_\sigma$. This choice presents a methodological advantage
without affecting the qualitative physical behavior of the model \cite{bible}. The unit of energy,
as customary, is set by $D_\uparrow=D=1$.
Within DMFT, the lattice model is map onto a quantum impurity problem 
subject to a self-consistency condition \cite{bible}.  The self-consistency condition is given by,
\begin{equation}
\label{eq:disordersc}
\mathcal{G}_{0\sigma}^{-1}(i\omega_n)=i\omega_n+\mu_\sigma-t_\sigma^2G_\sigma(i\omega_n)
\end{equation}
where $G_\sigma$ is the local Green's function of the  fermions with spin $\sigma$
and $\mathcal{G}_{0\sigma}$  are the bare Green's functions of the associated quantum 
impurity problem \cite{bible}.
The impurity problem can be solved using exact
diagonalization and quantum Monte Carlo techniques, whose implementation
has been already detailed elsewhere \cite{bible}.



The central quantities  in the DMFT method are the local Green functions, $G_\sigma(\omega_n)$, where
$\omega_n$ are the Matsubara frequencies. Here, since \tup $\neq$ \tdw 
we should generically have $ G_\uparrow \neq G_\downarrow $.
Though  the Matsubara Green's functions need
to be analytically continued to the real frequency axis 
to obtain the spectral function (density
of states), they nonetheless provide clear physical information. For instance, the density of states at the Fermi
energy, which signals metallic or insulating states, is directly given by  $-1/\pi {\rm Im}[G_\sigma(\omega_n
\rightarrow 0^+)]$. 

\begin{figure}
\centering
\includegraphics[width=7.7cm,angle=0]{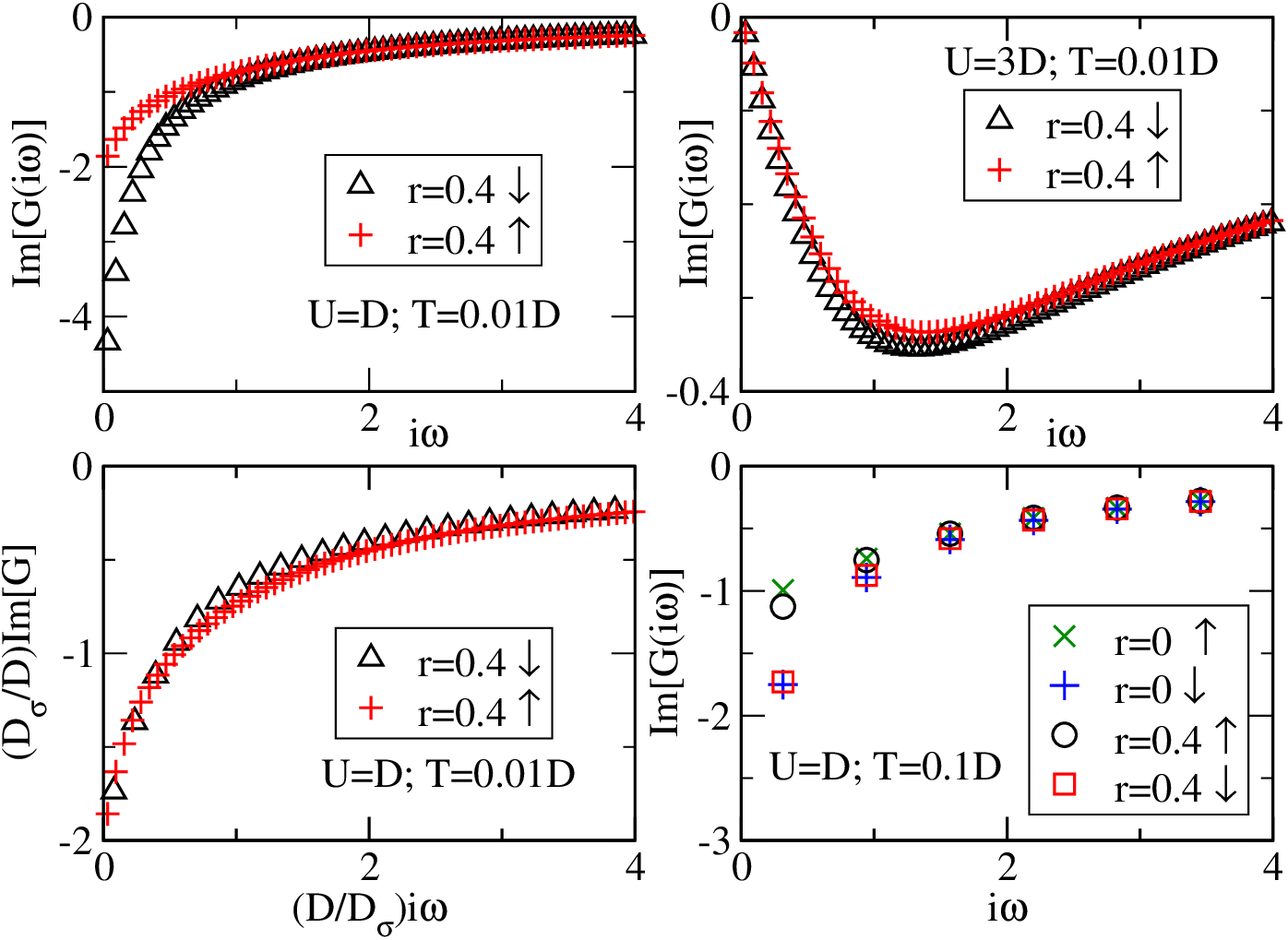}
\caption{\label{fig1}(Color online) Matsubara Green's functions of the AHM for $r$=0.4 and $T$=0.01$D$. 
The top left panel shows a metallic solution for $U$=$D$ and the right one show an insulating solution for $U$=3$D$.
The lower left panel shows the pinning condition of the two metallic solutions (Im[$G_\sigma](\omega_n \to 0) 
= 2/D_\sigma$ at $T=0$) \cite{bible}.
The lower right panel shows the same Green's functions at a higher temperature $T$=0.1$D >T_{coh}$.
The solution essentially coincides with the FKM solution ($r$=0) computed at the same value of $U$.
}
\end{figure}

At half-filling and $T=0$ it is well known that the FKM
has a Mott transition at a critical value  $U_C^{FKM}=D$ whereas the
HM is found to have
 two transitions at  $U_{c1}^{HM} \approx 2.4D$ and $U_{c2}^{HM} \approx 3D$.
The transition at $U_{c1}^{HM}$ is related to the closing of the Mott gap,
while that at $U_{c2}^{HM}$ stems  from the divergence of the carrier effective mass \cite{bible}.
Both metallic and insulating solutions coexist in the region
$U_{c1}^{HM}< U< U_{c2}^{HM}$.
Given that both the Hubbard and FKM have metallic and Mott insulating phases,  
we expect a  paramagnetic Mott metal-insulator transition  in the AHM as a function of
the repulsive interaction $U$  for all $r$. 
Our results for the Green's functions at the generic value $r=0.4$, for two values of the interaction $U=D$ and
3$D$ are shown  in the top left and right panels of Fig.\ref{fig1}.
The nonzero  y-axis intercepts of both, $G_\uparrow$ and $G_\downarrow$, in the data of the left top
panel indicate the presence of (different) metallic states in the two channels. 
On the other hand, as shown in the right top panel, at a higher value of the interaction $U$, 
the density of states vanishes, which signals an insulating state with the opening of a Mott gap for the up and down fermions.
Moreover,  as the gap is related to  the slope of
$G_\sigma(\omega_n)$ at low $\omega_n$, we see that the gap has a similar magnitude for the two species.   
Though this establishes the
existence of a Mott transition, we still need to understand 
how the Mott-transitions of the two end-cases (Hubbard and FKM)  merge as a function of $r$
and $U$.

We first analyze the metallic phase.
In the standard Hubbard model ($r$=1),  at low temperatures, one has  a metallic Fermi
liquid state  with quasiparticles which have a  renormalized mass $m^{*}$ given by the self-energy $\Sigma$ \cite{MAHAN}. 
In such a Fermi liquid state, within DMFT, at low frequencies
$\Sigma(\omega_n) \approx (1-1/Z) i \omega_n$, where $Z$ is the quasiparticle
residue \cite{bible}. 
Inputting the low frequency self energy  into the definition of the Green's function,  we obtain
\begin{equation} \label{eq:gfn}
G({\bf k},\omega_n) 
\approx 
{1 \over i\omega_n -\epsilon_{\bf k} -(1-{1\over Z}) i\omega_n} 
= {Z \over i\omega_n - Z \epsilon_{\bf k}}.
\end{equation}
As \eqref{eq:gfn} shows,   the effective mass renormalization factor $m^* \propto 1/Z$, and 
at the Mott transition $Z \to 0$ and $m^{*} \to \infty$.

In the asymmetric model it is not a priori evident how the interactions will renormalize each one of the
band masses. The results of the lower right panel of Fig.2 show the numerical results for $Z_\sigma$.
We find that the mass renormalization of the two species is different, with the lighter one being more 
strongly suppressed by the effect of interactions. However, they both vanish at the same critical point.
Thus, the ground-state of the system is an asymmetric Fermi liquid with two types of heavy quasiparticles, 
that beyond a critical value simultaneously localize into a Mott-Hubbard state.

\begin{figure}
\centering
\includegraphics[width=8cm]{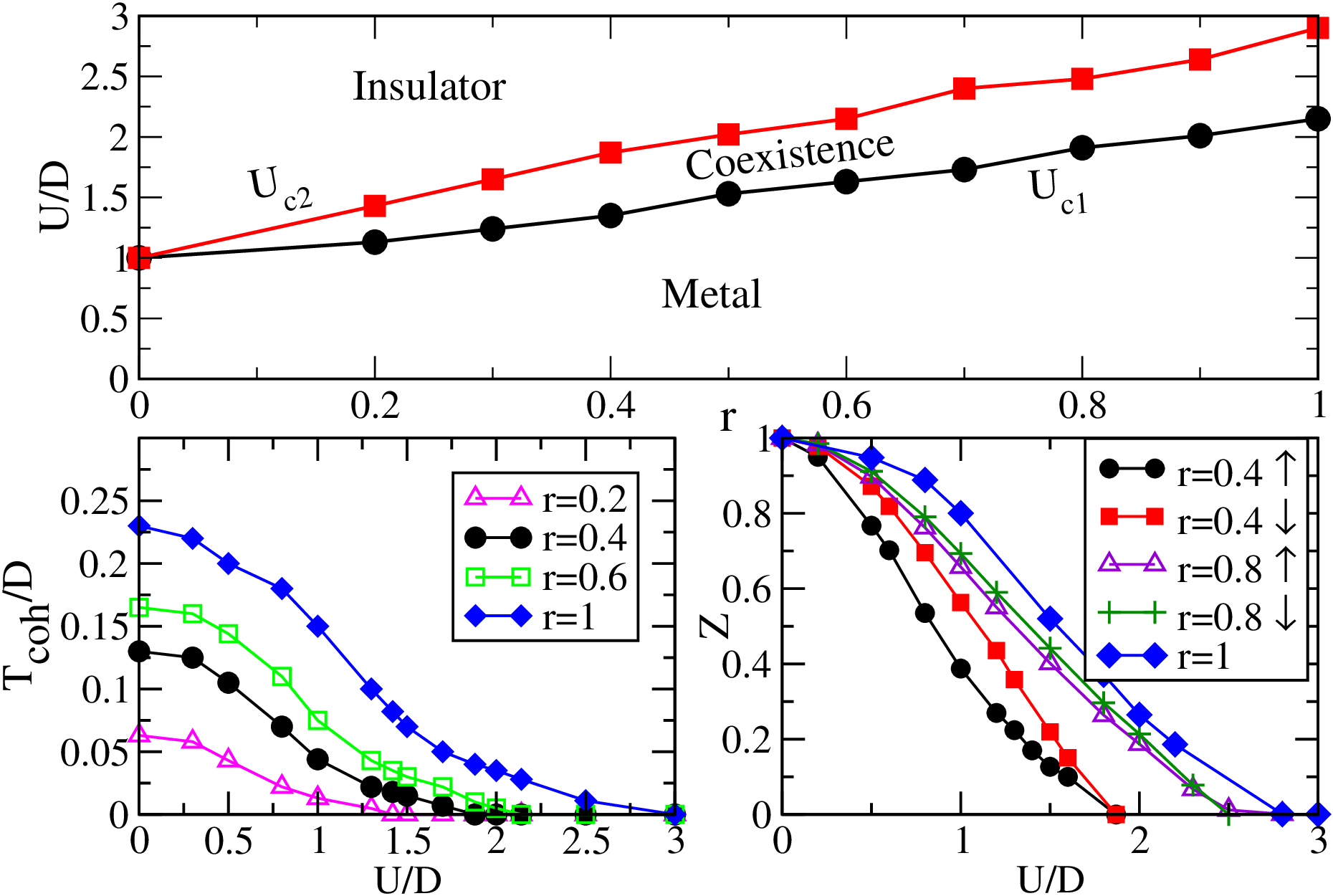}
\caption{\label{fig2}(Color online) The top panel shows the $T$ = 0 phase diagram
as function of $r$ and $U$ with the Mott transition boundaries $U_{c1}$ and $U_{c2}$. 
The lower left (right) panel shows the behavior of $T_{coh}$ ($Z_\sigma$) as a function of interaction $U$
for various values of the asymmetry factor $r$.  The size of the symbols are an estimate of the error bars.}
\end{figure}

As in the Hubbard case, the  Fermi liquid state seen in the AHM exists below a certain  coherence temperature
$T_{coh}$ and is characterized by a  specific heat  which is linear in temperature.  Above $T_{coh}$,  the 
thermal disorder is too strong for the quasiparticles to survive  and the good metallic behavior is gradually  lost. 
Our results for $T_{coh}$, obtained from the specific heat (to be discussed in detail later), are plotted in 
the lower left panel of Fig. \ref{fig2}.
We find that this temperature is the same for both species consistent with the single 
Kondo temperature of the associated impurity problem\cite{bible}.
As one increases the temperature above $T_{coh}$, the solution has a rapid crossover towards
a new bad  metallic state. Surprisingly, we find that this intermediate-temperature metallic state is
not arbitrary, but  essentially coincides with the non Fermi liquid  solution of the FKM obtained at the same
value of interaction $U$.
The corresponding numerical result is shown in the lower right panel of Fig.\ref{fig1}, where we compare the
solution obtained at  $T>T_{coh}$ with the FKM solution computed at the same value of $T$ and $U$.
We may interpret this rather striking result as if $r$ effectively renormalizes down to zero
as $T \sim T_{coh}$. 
Since in the FKM one of the fermionic species is itinerant while the other is fully localized (\tdw =0),
the observed behavior in the AHM amounts to an orbital selective crossover to localization of the down spin
fermions.
The AHM  can therefore be considered as a minimal model to realize an orbital selective transition, a concept
that has recently received a lot of attention in the context of strongly correlated fermions in condensed 
matter systems  \cite{vojta,luca}.

 The $T$=0  phase diagram of the model  as a function of asymmetry $r$
and interaction $U$ is shown in the top panel of  Fig.\ref{fig2}.  
As discussed before, at low $T$, the AHM qualitatively behaves as the Hubbard model and has two critical Mott transition 
boundaries, $U_{c1}$ and $U_{c2}$, that define a region of coexistent solutions \cite{bible}. The two critical lines
merge at the FKM end-point. This coexistence survives for a small range of temperatures  
resulting in a first order metal-insulator transition  line at
low temperatures which culminates in a second order critical point \cite{bible}.  
The calculation of this small critical temperature is technically demanding and is left for future work. 
Similarly to the Hubbard case,
we see in Fig.\ref{fig2} that
for all $r \neq 0$, $Z \to 0$  as $U \to U_{c2}$  leading to a  divergence of  the effective mass  at $U_{c2}$.

\begin{figure}
\centering
\includegraphics[width=7cm,angle=0]{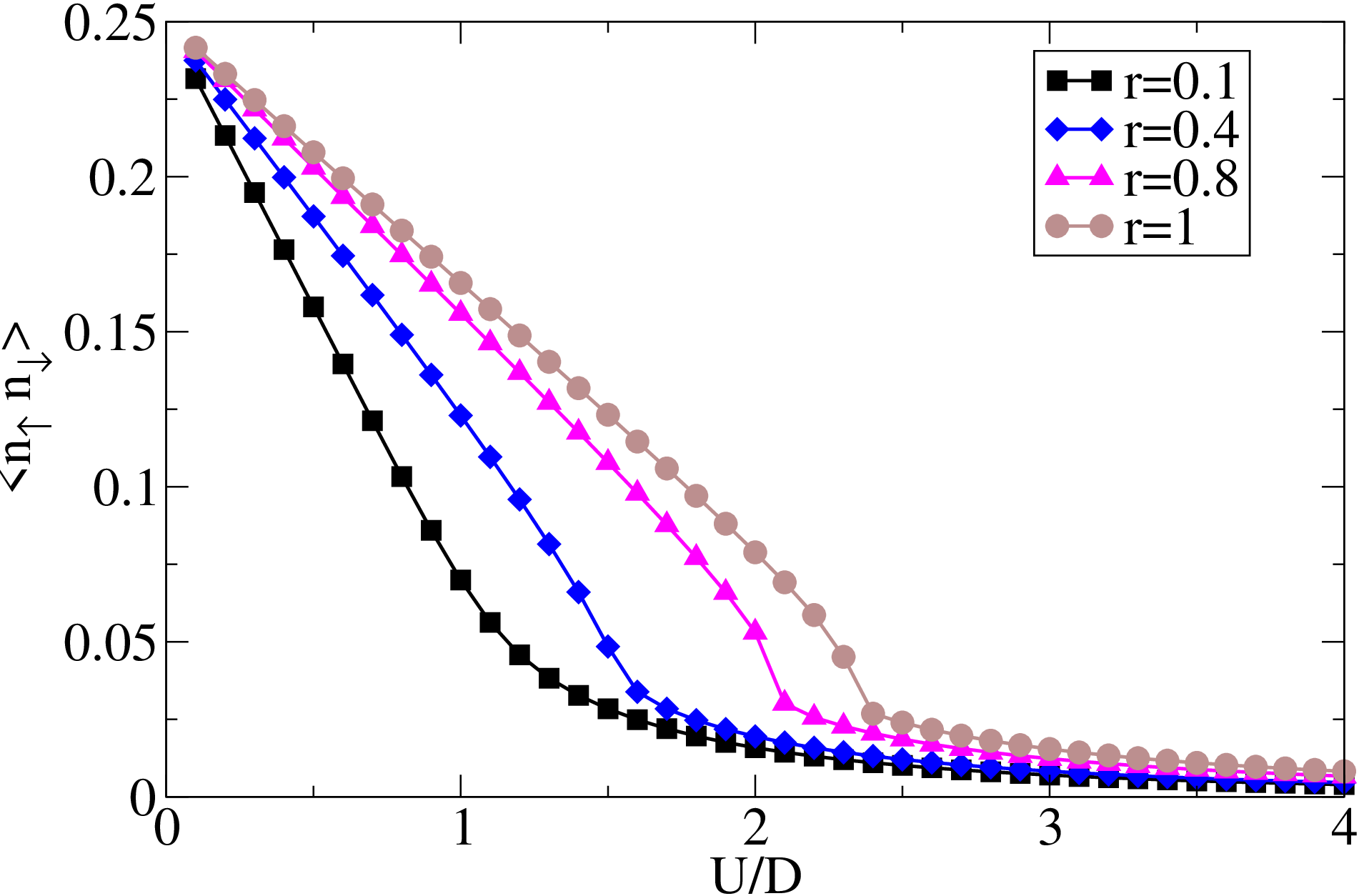}
\caption{\label{fig3}(Color online) Double occupancy as function of $U$ at $T$=0.025$D$, 
for different values or $r$=\ra. From top to bottom, 
the curves corresponds to decreasing values of $r$.}
\end{figure}

In order to establish a link between the behavior of the model and the
physical observables accessible in cold atom systems on optical lattices, we 
compute  the double occupation $\langle n_\uparrow n_\downarrow \rangle$, which can 
 be used as a 
thermometer \cite{jordensprl,esslingerprl}. In fact, this quantity is also closely related
to an order parameter for the Mott metal-insulator transition \cite{science,rck}.
The results are shown in Fig.\ref{fig3} for various  values of $U$ and $r$ at finite $T$=0.025$D$.
At larger values of $r$ there is a very sharp transition from the metal to the insulator state. The metal
is characterized by linear decrease of the double occupation with increasing interaction $U$, while 
in the insulator, at larger values of the interaction, the double occupation remains low and weakly
$U$-dependent. The rapid variation of $\langle n_\uparrow n_\downarrow \rangle$ with a sigmoidal shape between 
the two regimes
is also an indication of the finite-temperature region of coexistent solutions that we discussed before \cite{rck}.
At smaller values of $r$, when the system is more asymmetric, the double occupation rapidly decreases
due to the very small coherence scale and the reduced values of the critical $U$. For the lower $r$ values
the temperature of the calculation is actually above $T_{coh}$ ({\cal cf} Fig.\ref{fig2}), thus we observe a rather slow 
and wide crossover region for the evolution between the metal and the insulator. 

Other experimentally relevant quantities we compute  are  the  specific heat and
the  entropy which  are
required to  estimate the temperature of a cold atom system. In fact, these systems
are in the sub-millikelvin regime and since they do not dissipate energy their evolution 
is considered to occur essentially along isentropic lines \cite{agentropy}.
To understand the evolution of the specific heat and entropy, it is useful to start with
the known cases of the Hubbard and FKM at $T$=0. In a regular Fermi liquid the
entropy vanishes as $T \to 0$. However,
there are also two other cases to be considered, which are unusual as they
have unquenched entropy per site $S/N$=$\ln2$ at $T$=$0$ 
(we set the Boltzmann constant $k_B$=1). One is  
 the paramagnetic Mott-insulator, with one localized particle
per site, which only has the two spin-states as remanent degree of freedom.
The other case is the bad metallic state of the FKM, where the localized particles are disordered and have equal probability
of occupying or not each of the $N$ lattice sites, thus contributing the $\ln2$ to the entropy
at half-filling. The itinerant particles, in contrast, adjust to the potential created by the former
and form a band that contributes no entropy at $T=0$.  Consequently, starting
from the Fermi liquid state of the AHM,  we expect strong variations
of the specific heat for both, the
transition to the Mott insulator and the crossover to the orbital-selective bad metallic state.

\begin{figure}
\centering
\includegraphics[width=7cm,angle=0]{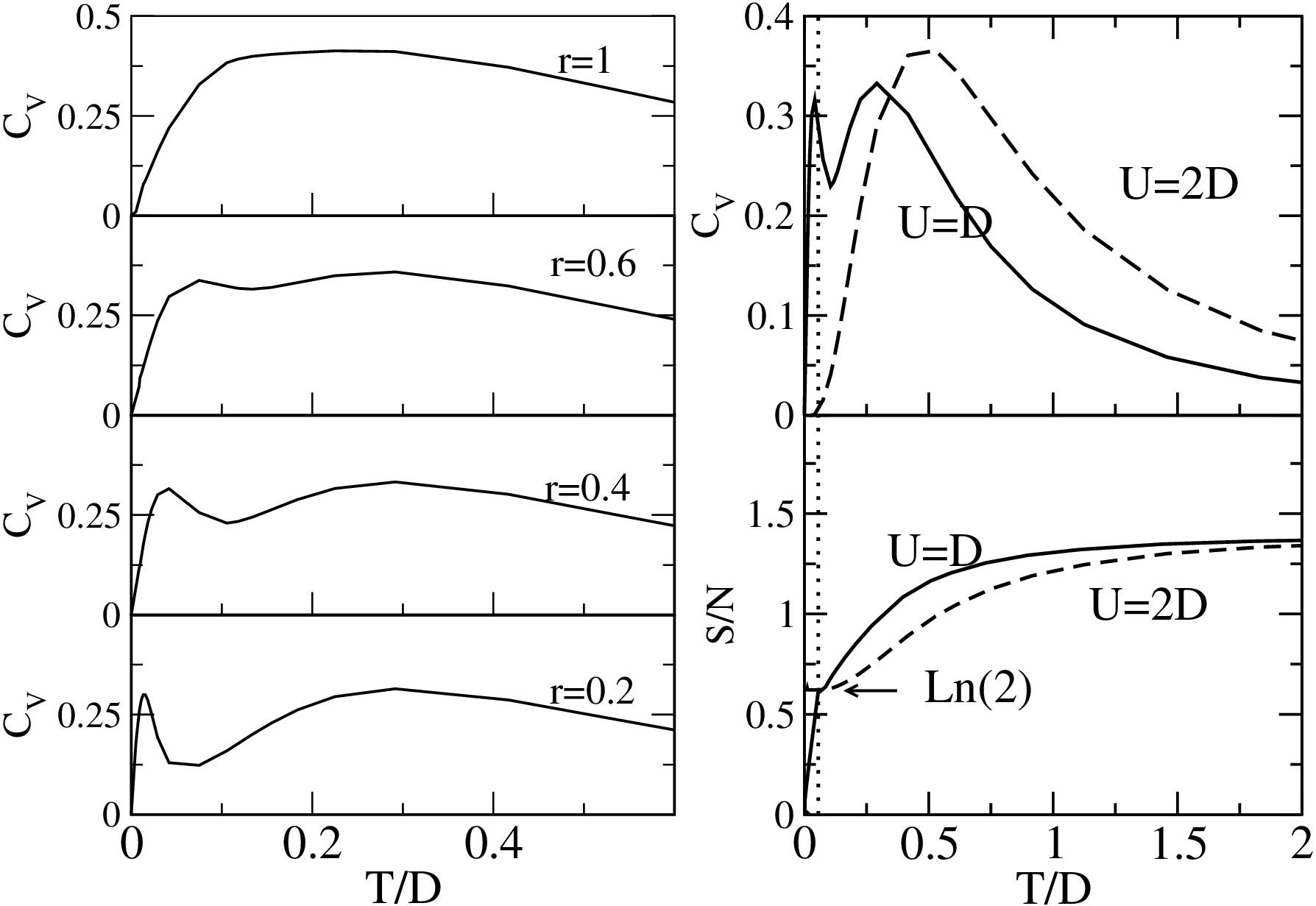}
\caption{\label{fig4}Left panels: Specific heat as function of $T$ for $U$=$D$ and different values of $r$. 
For $r$=1 the linear Fermi-liquid behavior is observed. As $r$ decreases, a two-peak structure is visible. The first peak 
moves to lower $T$ with decreasing $r$, revealing the reduction of $T_{coh}$ and the increase of the effective
mass (from the increase of the slope).Right panels: $C_v$ (top) and entropy (bottom) as a function of $T$ for $r$=0.4 and $U$=$D$
(metal) and $U$=2$D$ (Mott insulator).}
\end{figure}

Our results for the specific heat and entropy are plotted in Fig.\ref{fig4}.  The left side panels of Fig.\ref{fig4}
show the evolution of $C_v$ in the metallic phase for different values of $r$. Note that
for all $r\neq 0$ and low enough $T$ and $U$,  the system is a normal Fermi liquid with the specific heat
$C_v= \gamma T$  and $S \propto T$ with $\gamma$ proportional to the mean effective mass  
of the quasiparticles. As the temperature increases and reaches $T_{coh}$, the singlet Fermi-liquid metallic
state breaks down and the system crosses over to the orbitally selected 
bad-metal, which has FKM-like behavior, thus having an entropy of $\ln2$. This is clearly
seen in the lower right panel of Fig.\ref{fig4}, where after an initial linear increase, the
entropy attains a small plateau around $\ln2$. This plateau is concomitant with the 
first maximum observed in the $C_v(T)$ at a $T \sim T_{coh}$. Physically, it corresponds
to the orbital selection, i.e, the effective localization of the heavy-particle band.
As $T$ is further increased, the remaining degrees of freedom will eventually contribute
their $\ln2$ to the entropy, producing a second maximum in $C_v$ at 
the bare energy scale $T \sim t_\uparrow$=$D/2$. 
In the high-$T$ limit the  specific heat decreases and the entropy saturates to the maximal value  $\ln4$. 
For comparison, the behavior of the specific heat and the entropy of  a typical Mott
insulating  state at higher interaction $U=2D$ and $r=0.4$ is also shown in the right panel of Fig.{\ref{fig4}}. 
At low $T$,  $C_v(T)$ shows activated behavior  indicating  the 
presence of the Mott gap. We also observe, consistent with our previous
discussion, that the entropy starts from the unquenched value $\ln2$, and reaches $\ln4$
in the high-$T$ limit.

To conclude, we have studied the behavior of the asymmetric Hubbard model as a generic model
for  two species of fermionic cold atoms loaded onto an optical lattice. We determined the phase diagram
and found a Mott metal-insulator transition for all degrees of asymmetry.   The Mott transition scenario is the
same as that of the Hubbard model for all nonzero values of the asymmetry $r$.
Within the metallic state, 
we found and interesting temperature-driven orbital selective crossover.  
At low temperatures the AHM has the same qualitative behavior as the {\it symmetric} 
Hubbard model, however with two different heavy mass Fermi-liquids.
These two liquids have, nevertheless, a single coherence temperature.
Above a coherence temperature, one fermionic species localizes and  the system exhibits  
non-Fermi liquid behavior similar to that of 
  the Falicov-Kimball
model. We also computed the double occupation and some thermodynamical observables which may
permit the experimental identification of this remarkable physical behavior.  

E. A. W. is financially supported by the program ANR-09-RPDOC-019-01.


\end{document}